\documentstyle[multicol,aps,prb,amssymb,epsf]{revtex}

\begin{document}

\draft

\title{Interplay between disorder and intersubband collective
excitations in the two-dimensional electron gas}

\author{Stefano Luin,$^{1,2}$ Vittorio Pellegrini,$^1$ Fabio
Beltram,$^1$ Xavier Marcadet,$^2$ and Carlo Sirtori$^2$}

\address{$^1$Scuola Normale Superiore and INFM, I-56126 Pisa,
Italy\\ $^2$LCR Thales, Domaine de Corbeville, 91404 Orsay Cedex,
France}

\date{\today}

\maketitle

\begin{abstract}
Intersubband absorption in modulation-doped quantum wells is usually
appropriately described as a collective excitation of the confined
two-dimensional electron gas. At sufficiently low electron density and
low temperatures, however, the in-plane disorder potential is able to
damp the collective modes by mixing the intersubband charge-density
excitation with single-particle localized modes. Here we show
experimental evidence of this transition.  The results are analyzed
within the framework of the density functional theory and highlight
the impact of the interplay between disorder and the collective
response of the two-dimensional electron gas in semiconductor
heterostructures.
\end{abstract}
 
\begin{multicols}{2}
Intersubband (IS) excitations of two-dimensional electron gases
(2DEGs) confined in semiconductor heterostructures are among the most
important probes of electron-electron interactions\cite{1} and
represent an active area of research.\cite{2,3} Further motivation for
the experimental and theoretical study of IS transitions stems from
the growing number of opto-electronic devices whose operation is based
on these transitions.\cite{4,5} In the last decade several authors
have demonstrated that --particularly in the long-wavelength limit--
IS charge-density modes (the ones probed in optical absorption
experiments) are significantly shifted from the single-particle
transition energies by both direct (\emph{depolarization shift}) and
exchange-correlation (\emph{excitonic coupling}) terms of the Coulomb
interaction.\cite{6,7,8,9,10} The influence of these dynamic
contributions to the elementary excitation spectrum of 2DEGs has been
studied experimentally and theoretically modeled. Only in recent
times, however, attention has been given to the interplay between
disorder and many-body contributions.\cite{11,12,13} This interplay
raises several fundamental questions on the role of dephasing and
scattering in determining the collective response of electrons and has
direct implications on the optimization of IS-based devices.

Metzner \emph{et al.}\cite{11}\ and Ullrich \emph{et al.}\cite{13}\
addressed the influence on the IS linewidth of the various scattering
mechanisms and developed microscopic theories able to treat many-body
effects and disorder on equal footing. One of the main results of
these models is the precise description of how the in-plane disorder
destroys the coherence of the collective excitations and modifies the
IS linewidth and peak energy. In particular, Metzner \emph{et al.}\
emphasized that a transition from a regime where the intersubband
absorption is rather broad and originates from randomly distributed
localized single-particle excitations to a regime dominated by a sharp
and blue-shifted collective mode of the 2DEG can occur by varying the
electron density. This effect was explained as a mutual phase
adaptation of the localized oscillators interacting via long-range
Coulomb forces. Recently Yakimov \emph{et al.}\ reported experimental
evidence of such a transition in the case of holes in a
self-aggregated Si-Ge quantum dot matrix.\cite{15}
 
In this article we report a study of mid-infrared IS excitations in
2DEGs confined in modulation-doped narrow GaAs/Al$_{0.3}$Ga$_{0.7}$As
quantum wells (QWs) under the application of a perpendicular electric
field. The latter allows us to vary the carrier sheet density in the
range $10^9$--$10^{12}$~cm$^{-2}$ and explore regions where the
relative role of disorder and many-body effects are expected to be
markedly different without reaching the extreme localization limit
studied by Yakimov \emph{et al.}\ in Ref.\ \onlinecite{15}. We present
quantitative analysis of IS transitions based on a model that includes
both static and dynamic many-body effects. Our measurements reveal a
substantial softening of the IS excitation energy accompanied by an
abrupt increase of the intersubband absorption linewidth. This
behavior is observed at low temperatures and at electron densities
$\sim\!10^{11}$~cm$^{-2}$. We shall argue that at these densities a
transition occurs from a regime dominated by the collective response
of free electrons to another driven by the influence of localized
single-particle intersubband modes on this response. This is analogous
to the Landau damping mechanism occurring at finite in-plane
wavevector in a disorder-free 2DEG.\cite{6,12}
 
The two samples used for this study are modulation-doped single
GaAs/Al$_{0.3}$Ga$_{0.7}$As QWs (8.7-nm- and 7.5-nm-thick,
respectively). Doping in the Al$_{0.3}$Ga$_{0.7}$As layers was offset
by about 25~nm from the wells, and the growth was performed by
molecular beam epitaxy. At equilibrium samples have free-electron
concentrations close to $10^{12}$~cm$^{-2}$ and low-temperature
mobility above $10^5$~cm$^2$/Vs. Well thickness was chosen in order to
avoid population of the second subband and to maximize the impact of
fluctuations due to interface roughness, while preserving high values
of electron mobility.  Measured low-temperature IS transitions peak at
112.5~meV (8.7-nm-thick QW) and 130.5~meV (7.5-nm-thick QW). In order
to vary continuously the electron density in the samples, a metallic
gate was evaporated on the surface. In this way intersubband
absorption spectra were measured in the carrier-concentration range
$n\! \sim 2\!\times\!10^9$--$10^{12}$~cm$^{-2}$ and in the temperature
range 5--300 K. The 2DEG density was carefully determined by
Shubnikov-de Haas and Hall measurements as a function of gate voltage
$V_g$. For optical measurements, we fabricated $45^{\circ}$-edge
multipass waveguides. Absorption spectra were acquired with an
infrared Fourier transform interferometer in normal or step-scan modes
and using infrared light polarized perpendicularly to the QW layers
and propagating under the gate. In the step-scan case, the bias
voltage was square-wave modulated between -2V and the desired voltage,
at a frequency between 1 and 20 kHz. In both cases, background spectra
were collected at $V_g= -2$~V, at this bias the QWs were fully
depleted.
 
\begin{figure}
\centerline{\epsfxsize=3.2in
\epsffile{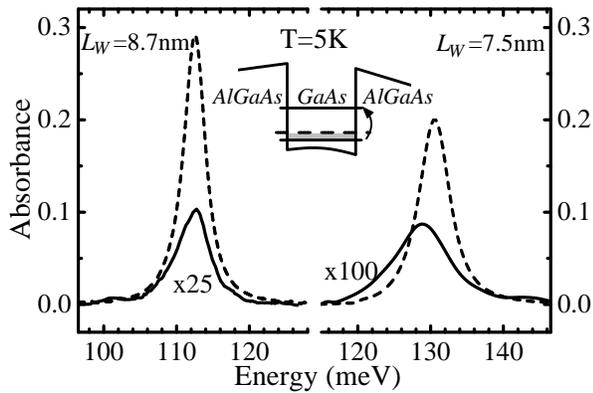}}
\caption{Low-temperature quantum well (QW) intersubband absorption
spectra at high (dashed line) and low (solid line) electron density
for the two structures studied.  Measurements were performed in a
2.6-mm-long and 370-$\mu$m-high optical waveguide.  Dashed lines
correspond to electron density $n\approx 8.4 \!\times\!
10^{11}$~cm$^{-2}$. Solid lines correspond to
$n\approx4\!\times\!10^{10}$~cm$^{-2}$ (left panel), and to $n \approx
1 \!\times\!10^{10}$~cm$^{-2}$ (right panel). Inset: Calculated band
diagram for the 8.7~nm QW with no applied bias. Horizontal solid lines
correspond to the energy of the bottom of the first two subbands;
horizontal dashed line is the chemical potential.} \label{fig1}
\end{figure}

\begin{figure}
\centerline{\epsfxsize=3.2in
\epsffile{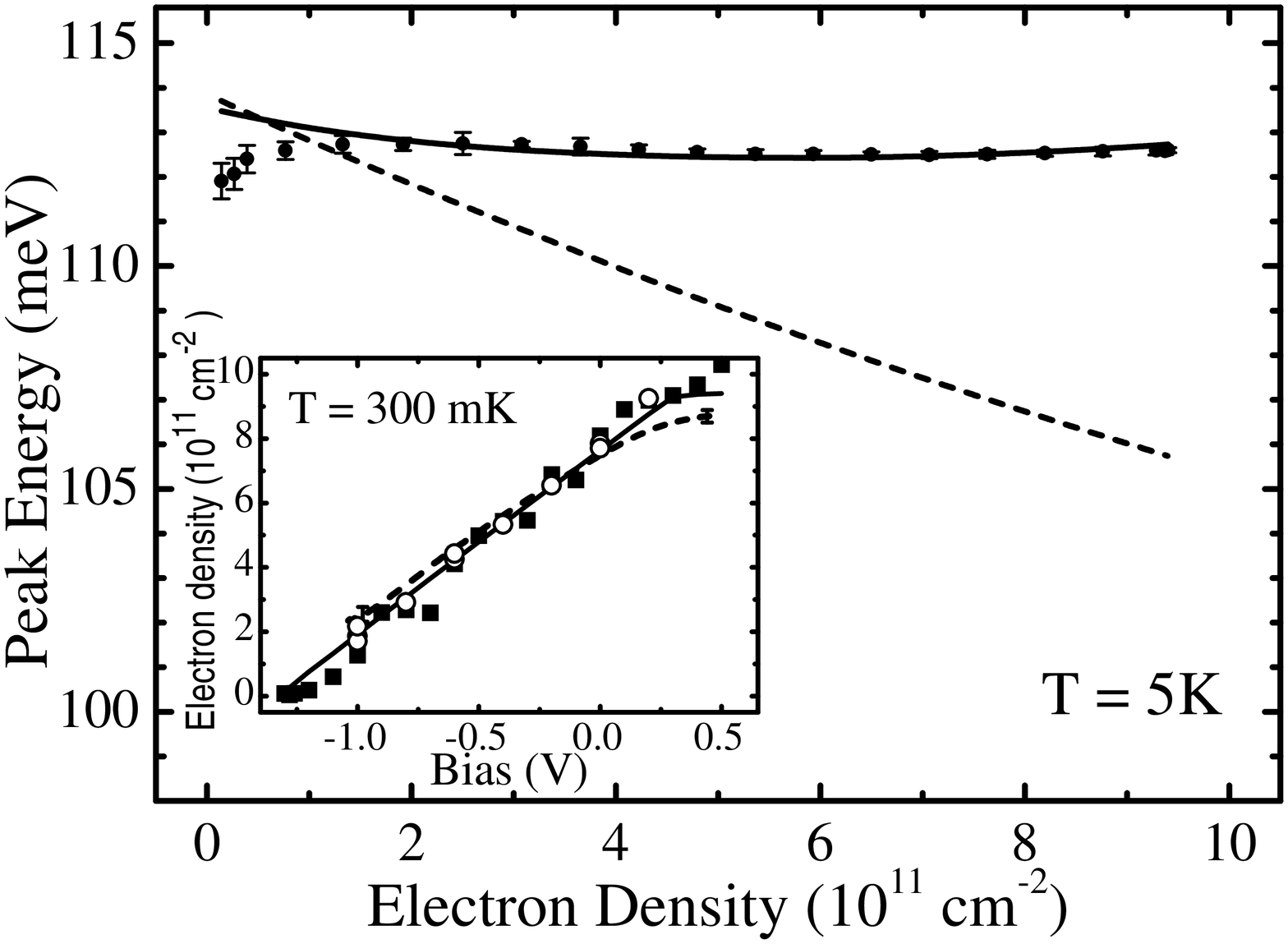}}
\caption{Low-temperature intersubband absorption peak energies as a
function of carrier sheet density for the sample with 8.7~nm well
width; experimental data are shown as dots with error bars. Solid
line: calculated peak energy including many-body effects within the
adiabatic local density approximation of the density functional
theory. Dashed line is the Hartree approximation. Inset: electron
sheet density as a function of applied bias as deduced from Hall
(dotted line), Shubnikov-de Haas (open circles), and optical
measurements (squares). Solid line is the result of the theoretical
model.}
\label{fig2}
\end{figure}

Figure \ref{fig1} shows representative absorption spectra for the
8.7-nm- (left panel) and 7.5-nm-QW structures (right panel) out of a
large set taken at many gate-bias values and different temperatures.
Dashed lines refer to electron density $n\approx 8.4 \!\times\!
10^{11}$~cm$^{-2}$ ($V_g=0.1$~V and 0.4~V in the 8.7 and 7.5~nm QWs,
respectively), solid lines to $n\approx 4 \!\times\!
10^{10}$~cm$^{-2}$ ($V_g= -1.2$~V, left panel) and $n\approx 1
\!\times\! 10^{10}$~cm$^{-2}$ ($V_g= -1.1$~V, right panel). Data
correspond to transitions between the first two subbands in the well,
as depicted in the inset of Fig.~\ref{fig1}. At the higher electron
densities, spectra have a lorentzian lineshape with a full width at
half maximum (FWHM) of 3.5 and 4.5~meV. Surprisingly, the measured
peak-energy shift at the different biases in Fig.~\ref{fig1} is very
small, despite the very different electric fields (up to
$8\!\times\!10^4$~V/cm at the lower carrier densities) and carrier
concentrations. Indeed, a single-particle description for the IS
energy that includes only the static Hartree term (and therefore takes
into account exclusively the appropriate band-bending of the biased
heterostructures) does lead to the prediction of a significant energy
shift (see Fig.~\ref{fig2}, dashed line for the case of the 8.7~nm
QW). The observed weak dependence of IS transition energy on electron
density is the result of a cancellation effect between Stark shift and
dynamic many-body contributions.\cite{7} This is clearly shown in
Fig.~\ref{fig2} where measured IS peak energies for the 8.7~nm
structure (dots) are plotted as a function of carrier sheet density
together with results of a theoretical model (solid line) that
includes many-body contributions within the adiabatic local density
approximation (LDA) of the density functional theory, and therefore
explicitly takes into account the effects of depolarization and
excitonic shifts. The IS peak energy was calculated by evaluating the
maximum of:
\begin{equation}
{{\cal{R}}\!e} \tilde{\sigma}_{zz}(\omega) \propto {{\cal{I}}\!m}
\frac{\omega {G}_{12}(\omega)}{1 + \left( \alpha(\omega) - \beta
\right) {G}_{12}(\omega)}\,,
\label{eq1}
\end{equation}
where $\tilde{\sigma}_{zz}(\omega)$ is the $zz$ component of the
frequency dependent conductivity tensor, and $G_{12}$ is proportional
to the response function of a non-interacting 2DEG including the
specific energy dispersions of the two individual subbands
considered. These are evaluated extending the calculations reported in
Ref.\ \onlinecite{16} and considering the band bending in the entire
heterostructure (see also Ref.\ \onlinecite{7}); in all the theory
different non-parabolic dispersions in every layer were
considered.\cite{16} In~(\ref{eq1}) the depolarization contribution is
linked to
\begin{equation}
\alpha(\omega) = 2e^2 n \int \frac{{\textrm{d}}
z'}{\varepsilon(z',\omega)} \left[ \int_0^{z'} {\textrm{d}} z \xi_2
(z) \xi_1(z)\right]^2 \frac{ 1 }{ {\mathcal{E}}_{12}}\,,
\end{equation}
where $\varepsilon(z,\omega)$ is the layer-dependent dielectric
function, whose frequency dependence is estimated like in Ref.\
\onlinecite{9}, $\xi_{1,2}$ are the wave-functions along the growth
direction and ${\mathcal{E}}_{12}$ is the difference between the
minima of the two subbands, as calculated from a Schr\"odinger type
equation derived in the LDA formulation of density functional
theory. The excitonic contribution is linked to
\begin{equation}
\beta = -2n\int {\textrm{d}} z \xi_2(z)^2 \xi_1(z)^2
\frac{{\textrm{d}} v_{xc}}{{\textrm{d}} N}
\frac{1}{{\mathcal{E}}_{12}}\,,
\end{equation}
where $v_{xc}(N)$ is the exchange-correlation potential\cite{17} as a
function of the position-dependent three-dimensional electron density
$N$ determined self-consistently.

The inclusion of all these effects was necessary to obtain the
successful description of the experimental results in the broad range
of carrier density that is shown in Fig.~\ref{fig2}. In particular,
the accuracy of our model in determining the 2DEG density is shown in
the inset of Fig.~\ref{fig2}, where the carrier density estimated by
Hall measurements (short-dash line), Shubnikov-de Haas oscillations
(empty circles), and area under the absorbance peak (with oscillator
strength calculated as in Ref.\ \onlinecite{18}) is reported as a
function of $V_g$ together with the theoretical dependence as derived
within LDA (solid line). Figure~\ref{fig2} shows also an excellent
agreement between experimental and theoretical results for IS
transition energies in a wide range of electron density (similar
results were obtained for the second sample). This agreement however
is not present at low densities where the theory is not able to
reproduce the softening of the IS transition energy. This behavior
signals a change in the nature of the intersubband transition and
originates from an interplay between many-body effects and
disorder. This is the focus of this paper and in what follows we shall
discuss the manifestations of this transition and its physical
meaning.
 
\begin{figure}
\centerline{\epsfxsize=3.0in
\epsffile{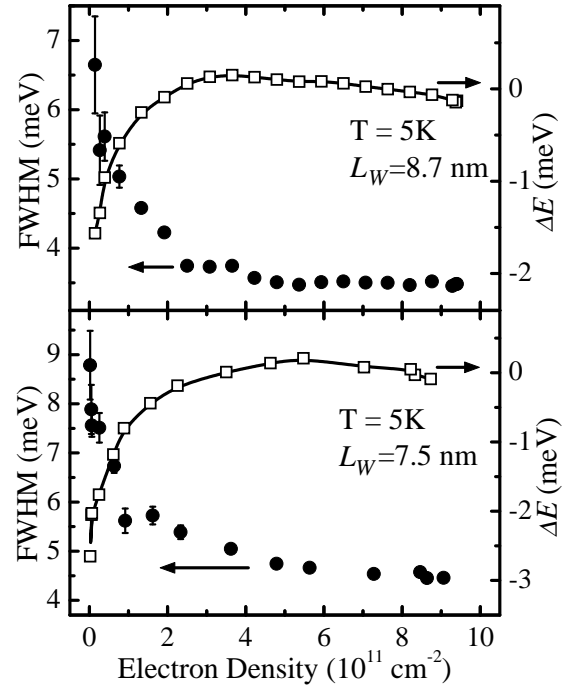}}
\caption{Full width at half maximum (FWHM, solid circles) of the
intersubband absorption peak and deviation of measured peak position
from theoretical values ($\Delta E$, open squares) as a function of
electron density.} \label{fig3}
\end{figure}

In order to analyze the low-electron density behavior of IS
transitions, it is useful to refer to Fig.~\ref{fig3} where the
deviation of the experimental data from the calculated values is
reported for both samples (open squares, right scale) at T $=5$~K. At
high density this deviation is $\approx\! 0$~meV for both samples, but
for $n < 3\!\times\!10^{11}$~cm$^{-2}$ (8.7~nm sample, upper panel)
and $n < 4\!\times\!10^{11}$~cm$^{-2}$ (7.5~nm sample, lower panel) we
observe a notable drop in energy of about 2~meV and 3.5~meV,
respectively. This evolution was monitored down to very low densities
($1.4\!\times\!10^{10}$~cm$^{-2}$ and $2\!\times\!10^9$~cm$^{-2}$,
respectively).  The pronounced decrease of the intersubband absorption
spectral weight hindered measurements at even lower
densities. Remarkably, the softening of the IS transition energies
occurs together with a rather sudden increase of the corresponding
FWHM values. This is also shown in Fig.~\ref{fig3} (solid circles,
left scale) where experimental FWHMs are plotted as a function of
electron density. In the high density regime the FWHM stays close to a quite
small value (around 4~meV). This type of behavior is consistent
with calculations that include many-body
contributions. In fact it has been shown, and confirmed by our
calculation, that these terms bring to its intrinsic homogeneous
value\cite{note} the absorption-peak linewidth drastically
reducing it from the values calculated taking into account the
different and non-parabolic dispersions of the subbands, that
increase with $n$.\cite{16,19} Both the energy position and FWHM
trend, therefore, indicate that in the high-density regions absorption
originates from purely collective intersubband excitation (charge
density excitation -- CDE).
 
\begin{figure}
\centerline{\epsfxsize=3.2in
\epsffile{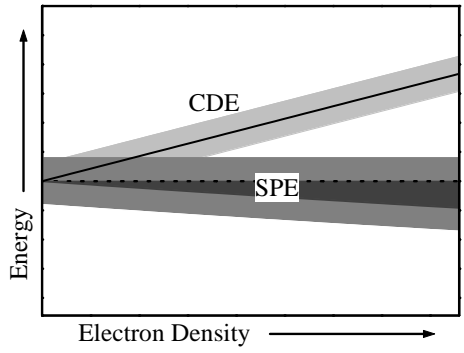}}
\caption{Schematic representation of charge-density excitation (CDE)
energy (solid line) and intersubband single particle excitation (SPE)
continuum (gray and dark gray areas) versus electron sheet
density. The light gray area represents the intrinsic CDE
broadening. The darker gray area shows the broadening originating from
subband dispersions, the gray region the inhomogeneous broadening.}
\label{fig4}
\end{figure}

The observed behavior at the lower densities signals that a change
takes place in the collective response of the 2DEG. This transition
occurs when the Coulomb interaction between electrons is not strong
enough compared to disorder and cannot concentrate the whole
oscillator strength in a narrow CDE mode.\cite{11} In this regime the
IS transition incorporates contributions of weakly-interacting
randomly localized excitations, and this leads to the observed peak
broadening and its shift towards lower energies. This evolution is
schematically illustrated in Fig.~\ref{fig4}, where the trends in the
CDE energy (black line) and in the single particle excitation (SPE)
continuum (darker gray areas) are shown as a function of density. The
light gray area represents the intrinsic CDE broadening. The darkest
gray area includes the non-parabolicity contributions. The dominant
source of broadening for single-particle excitations (SPEs), however,
is inhomogeneous in nature and gives rise to the overall SPE
broadening shown in Fig.~\ref{fig4}. The observed transition takes
place when the broadened CDE and SPE excitations merge. At this
density new relaxation channels of the intersubband CDE into localized
single-particle modes open, a phenomenon similar to conventional
Landau damping at finite in-plane wavevectors.\cite{6,12} At even
lower densities intersubband spectra are characterized by localized
SPEs. This second regime was explored by Yakimov \emph{et
al.}\cite{15} thanks to the large fluctuations provided by the
self-aggregated quantum dots in the samples they studied.  This
overall behavior was theoretically discussed in Ref.\
\onlinecite{11}. It can be noted that the transition region here
discussed occurs at higher density for the narrower well, where indeed
the electron states are more sensitive to well-width
fluctuations. Furthermore, we observed that these changes in energy
and linewidth at low densities display a marked temperature dependence
and were not detected at temperatures larger than 60~K (data not
shown). This suggests an intriguing thermally-induced modification of
the collective response of the system. It must be noted in fact that,
contrary to this behavior, the many-body contributions to the peak
position persisted at all temperatures studied.

However, in order to further corroborate our interpretation, we
developed a simple model based on the Drude approach for the response
function of a QW characterized by a gaussian-broadened single-particle
intersubband transition with a phenomenological inhomogeneous
broadening superimposed to a costant homogeneous linewidth.\cite{20}
Results for the deviation of the peak position from the
depolarization-shifted energy and for the FWHM, as obtained by a
lorentzian fit of the calculated response function, are in agreement
with the data reported in Fig.~\ref{fig3}. Within this simplified
model, the transition between the purely collective excitation and the
intermediate regime discussed above occurs when the depolarization
shift is comparable to the inhomogeneous broadening consistently with
our data. Also, the softening of the transition energy occurs because
the CDE peak, located in the high-energy end of the broadened
single-particle excitation spectrum, possesses all the oscillator
strength at high density.

In conclusion, we have experimentally and theoretically studied
intersubband transitions in modulation-doped narrow GaAs quantum wells
in a wide range of electron densities. We have reported evidence of
the damping of the CDE due to coupling with single-particle localized
modes as the density is lowered. This transition is a direct
manifestation of the influence of the interplay between disorder and
many-body effects in determining the nature of IS excitations of the
2DEG in semiconductor heterostructures.

We thank A. Tredicucci, G. Vignale and B. Vinter for useful
discussion. The work at Scuola Normale was partially supported by
MURST.

\end{multicols}

\end{document}